\newcommand{\p}{\partial}
\newcommand{\pq}{\partial_{q}}
\newcommand{\tht}{\theta}

\newcommand{\Wr}{\mathrm{Wrd}}

\newcommand{\af}{\alpha}
\newcommand{\ta}{\tau}
\newcommand{\et}{\eta}

\newcommand{\dt}{\delta}

\newtheorem{prop}{Proposition}

\newtheorem{lemm}{Lemma}
\newtheorem{eg}{Example}
\newtheorem{rmk}{Remark}

\documentclass[12pt]{myiopart}

\eqnobysec

\begin{document}

\title[The $q$-mKP hierarchy with sources, Wronskian solutions and solitons]{The $q$-deformed mKP hierarchy with self-consistent sources, Wronskian solutions and solitons}

\author{Runliang Lin$^{1,2}$, Hua Peng$^1$ and Manuel Ma\~{n}as$^2$}

\address{$^1$ Department of Mathematical Sciences, Tsinghua University, Beijing 100084, P.R. China}
\address{$^2$ Departamento de F\'{\i}sica Te\'{o}rica II, Universidad Complutense, 28040 Madrid, Spain}
\ead{\mailto{rlin@math.tsinghua.edu.cn}, \mailto{pengh05@126.com}, \mailto{manuel.manas@fis.ucm.es}}

\begin{abstract}
Based on the eigenfunction symmetry constraint of the $q$-deformed
modified KP hierarchy, a $q$-deformed mKP
hierarchy with self-consistent sources ($q$-mKPHSCSs) is constructed.
The $q$-mKPHSCSs contains two types of $q$-deformed mKP equation with self-consistent sources.
By combination of the dressing method and the method of variation of constants, a
generalized dressing approach is proposed to solve the
$q$-deformed KP hierarchy with self-consistent sources ($q$-KPHSCSs). Using the gauge
transformation between the $q$-KPHSCSs and the $q$-mKPHSCSs, the
$q$-deformed Wronskian solutions for the $q$-KPHSCSs and the $q$-mKPHSCSs are obtained.
The one-soliton solutions for the $q$-deformed KP (mKP) equation with a source are given explicitly.

\end{abstract}

\pacno{02.30.Ik}
\maketitle

\section{Introduction}
In recent years, the $q$-deformed integrable systems attracted many
interests both in mathematics and in physics
(see, e.g., 
[1-13]).
The $q$-deformation is performed by using the $q$-derivative
$\partial_q$ to take the place of ordinary derivative $\partial_x$
in the classical systems, where $q$ is a parameter, and the
$q$-deformed integrable systems recover the classical ones as
$q\rightarrow 1$.
The $q$-deformed integrable systems usually inherit some integrable structures
from the classical integrable systems. Take the $q$-deformed KP hierarchy as an example,
its $\tau$-function, bi-Hamiltonian structure and additional symmetries have already been reported (see \cite{Iliev1998,Iliev2000,Tu1999,HLC} and the references therein).

Multi-component generalization of an integrable model is a very
important subject
(see, e.g., 
[14-16]).
For example, the multi-component KP (mcKP) hierarchy given in
\cite{MR638807} contains many physically relevant nonlinear
integrable systems, such as Davey-Stewartson equation,
two-dimensional Toda lattice and three-wave resonant interaction
ones. Another type of coupled integrable systems is the soliton
equation with self-consistent sources, which has many physical
applications and can be obtained by coupling some suitable
differential equations to the original soliton equation
[17-23].
Recently, a systematical procedure is proposed to construct a
KP hierarchy with self-consistent sources and its Lax representation \cite{LZL,dress}.
This idea can be used to the $q$-deformed case, i.e.,
by introducing a new vector filed
$\partial_{\tau_k}$ as a linear combination of all vector fields
$\partial_{t_n}$ in the $q$-deformed KP hierarchy, then one can
get a new Lax type equation which consists of the
$\tau_k$-flow and the evolutions of wave functions. Under the
evolutions of wave functions, the commutativity of
$\partial_{\tau_k}$-flow and $\partial_{t_n}$-flows gives rise to a
$q$-KP hierarchy with self-consistent sources ($q$-KPHSCSs) \cite{Lin:ext-qKP}.
The $q$-KPHSCSs
consists of $t_n$-flow, $\tau_k$-flow, and $t_n$-evolutions of the
eigenfunctions and adjoint eigenfunctions. This $q$-KPHSCSs contains
two types of $q$-deformed KP equation with self-consistent sources
(1{\it{st}}-$q$-KPSCS and 2{\it{nd}}-$q$-KPSCS), and the two kinds of reductions of the $q$-KPHSCSs give
the $q$-deformed Gelfand-Dickey hierarchy with self-consistent
sources and the constrained $q$-deformed KP hierarchy, respectively,
which are some (1+1)-dimensional $q$-deformed soliton equation with
self-consistent sources \cite{Lin:ext-qKP}.

The dressing method is an important tool for solving the soliton
hierarchy \cite{Dickey}.  However, this method can not be applied
directly to solve the hierarchy with self-consistent sources.
In this paper, with the combination of dressing method and the
method of variation of constants, a generalized dressing method for
solving the $q$-KPHSCSs is proposed. In this way, we can get some solutions of the
$q$-KPHSCSs in a unified and simple procedure. As a special case,
the $N$-soliton solutions of the two types of $q$-KPSCS's are
obtained simultaneously.

To our knowledge, compared to the study on the $q$-deformed KP hierarchy,
there are less results on the {\it $q$-deformed} modified KP hierarchy ($q$-mKPH).
Takasaki studied the $q$-mKPH and its quasi-classical limit
by considering the $q$-analogue of the tau function of the modified KP hierarhcy \cite{Takasaki-2005}.
As another part of this paper, we will present the $q$-mKPH explicitly,
 and then construct an $q$-deformed mKP
hierarchy with self-consistent sources ($q$-mKPHSCSs) on the base of eigenfunction
symmetry constraint. The $q$-mKPHSCSs provides a unified way to
construct two types of $q$-deformed modified KP equation with self-consistent sources (1{\it{st}}-$q$-mKPSCS and 2{\it{nd}}-$q$-mKPSCS).
Then a gauge transformation between the $q$-KPHSCSs and the $q$-mKPHSCSs is
presented. Since the Wronskian solutions to the $q$-KPHSCSs have been obtained by a generalized
dressing approach in former part of this paper, the gauge transformation enables us to get the
explicit formulation of $q$-deformed Wronskian solutions for
the $q$-mKPHSCSs.
It should be noticed that a general setting of ``pseudo-differential"
operators  on regular time scales has been proposed to construct some integrable systems \cite{GGS2005,BSS2008},
  where the $q$-differential operator is just a particular case.

This paper will be organized as follows. In Sec. 2, we briefly
recall how to construct the $q$-KPHSCSs and its Lax representation,
and it is shown that the $q$-KPHSCSs includes two types of the $q$-KPSCS.
In Sec. 3, a generalized dressing method for
the $q$-KPHSCSs will be proposed. In Sec. 4, a
$q$-deformed mKP hierarchy with self-consistent sources ($q$-mKPHSCSs) will be constructed,
and which includes two types of $q$-deformed mKP equation with self-consistent sources.
In Sec. 5, the gauge
transformation between the $q$-KPHSCSs and the $q$-mKPHSCSs is
established. In Sec. 6, the one-soliton solutions of the $q$-deformed KP (mKP) equation with a source
are obtained. In the last section, some conclusions and remarks will be given.

\section{The $q$-KP hierarchy with self-consistent sources ($q$-KPHSCSs)}
First, we introduce some useful formula for $q$-KP hierarchy. We
denote the $q$-shift operator and the $q$-difference operator by
$\theta$ and $\pq$, respectively, where $q$ is a parameter. These
operators act on a function  $f(x)$ ($x\in \bf{R}$) as
    $$ \theta (f(x))=f(qx), \qquad\pq (f(x))=\frac{f(qx)-f(x)}{x(q-1)}.$$
In this paper, we use $P(f)$ to denote an action of a difference
operator $P$ on the function $f$, while $P\circ f=Pf$ means the
multiplication of a difference operator $P$ and a zero-{\it th} order difference
operator $f$, e.g., $\partial_q f=(\partial_q(f))+\theta(f)
\partial_q$.

Let $\pq^{-1}$ denote the formal inverse of $\pq$.
In general, the following $q$-deformed Leibnitz rule holds
\begin{equation}
 \label{eqn:def1}
    \partial_q^n f=\sum_{k\ge0}\left(\begin{array}{c}{n}\\{k}\end{array}\right)_q\theta^{n-k}(\partial_q^kf)\partial_q^{n-k},\qquad n\in
     {\bf{Z}},
\end{equation}
where the $q$-number and the $q$-binomial are defined by
\begin{equation*}
\fl
(n)_q=\frac{q^n-1}{q-1},
\qquad
   \left(\begin{array}{c}{n}\\{k}\end{array}\right)_q=\frac{(n)_q(n-1)_q\cdots(n-k+1)_q}{(1)_q(2)_q\cdots(k)_q},\qquad
   \left(\begin{array}{c}{n}\\{0}\end{array}\right)_q=1.
\end{equation*}
For a $q$-pseudo-differential operator ($q$-PDO) of the form
$P=\sum\limits^n_{i=-\infty}p_i\pq^i$, we decompose $P$ into the
differential part $P_{+}=\sum\limits^n_{i=0}p_i\pq^i$ and the
integral part $P_{-}=\sum\limits_{i\leq-1}p_i\pq^i$. And the
conjugate operation ``$*$" for
 $P=\sum\limits^n_{i=-\infty}p_i\pq^i$ is defined by $P^*=\sum\limits^n_{i=-\infty}(\pq^*)^ip_i$ with
 $$\pq^*=-\pq\tht^{-1}=-\frac1q\partial_{\frac1q}, \qquad (\pq^{-1})^*=(\pq^*)^{-1}=-\tht\pq^{-1}.$$
The $q$-exponent $e_q(x)$ is defined as
    $$e_q(x)=\exp(\sum^{\infty}_{k=1}\frac{(1-q)^k}{k(1-q^k)}x^k).$$
Then it is easy to prove that $\pq^k (e_q(xz))=z^ke_q(xz),$
$k\in\bf{Z}$.

The Lax equation of the $q$-KP hierarchy ($q$-KPH) is given by (see, e.g., \cite{Iliev1998})
\begin{equation}
  \label{eqn:dKP-LaxEqn}
  L_{t_n}=[B_n,L]\equiv B_nL-LB_n,
\end{equation}
where  $L = \pq + u_0 + u_1\pq^{-1}+u_2\pq^{-2}+\cdots $,
$B_n=L^n_{+}$ stands for the differential part of $L^{n}$.
For any fixed $k\in{\bf {N}}$, by introducting a new variable $\tau_k$
whose vector field is
\begin{equation*}
    \partial_{\tau_k}=\partial_{t_k}-\sum_{i=1}^N\sum_{s\ge0}\zeta_i^{-s-1}\partial_{t_s},
\end{equation*}
where $\zeta_i$'s are arbitrary distinct non-zero
parameters,
then a $q$-deformed KP
hierarchy with self-consistent sources ($q$-KPHSCSs) can be constructed as following \cite{Lin:ext-qKP}
\numparts
  \label{eqns:exdKP-LaxEqn}
  \begin{eqnarray}
    L_{\ta_k}&=[B_k+\sum_{i=1}^N\phi_i\pq^{-1}\psi_i,L], \label{eqn:exdKP-LaxEqna}\\
    L_{t_n}&=[B_n,L],  \qquad \forall n\neq k, \label{eqn:exdKP-LaxEqnb}\\
    \phi_{i,t_n}&=B_n(\phi_i),\qquad
    \psi_{i,t_n}=-B_n^*(\psi_i),\qquad i=1,\cdots,N. \label{eqn:exdKP-LaxEqnc}
  \end{eqnarray}
\endnumparts

The following proposition is proved in \cite{Lin:ext-qKP}.

\begin{prop}
  The commutativity of (\ref{eqn:exdKP-LaxEqna}) and (\ref{eqn:exdKP-LaxEqnb}) under (\ref{eqn:exdKP-LaxEqnc})
   gives rise to the following zero-curvature representation for
   $q$-KPHSCSs (\ref{eqns:exdKP-LaxEqn})
\numparts
    \label{eqns:extdkp}
    \begin{eqnarray}
      B_{n,\tau_{k}}&-(B_{k}+\sum\limits_{i=1}^{N}\phi_{i}\pq^{-1}\psi_{i})_{t_{n}}+[B_{n},B_{k}
      +\sum\limits_{i=1}^{N}\phi_{i}\pq^{-1}\psi_{i}]=0,\label{eqn:extdkpa}\\
      \phi_{i,t_n}&=B_n(\phi_i),\qquad
      \psi_{i,t_n}=-B_n^*(\psi_i),\qquad i=1,2,\cdots,N,\label{eqn:extdkpb}
    \end{eqnarray}
\endnumparts
  with the  Lax representation given by
\begin{equation}
  \label{eqn:exdKP-LaxPair}
  \Psi_{t_n}=B_n(\Psi),\quad \Psi_{\ta_k}=(B_k+\sum_{i=1}^N\phi_i\pq^{-1}\psi_i)(\Psi).
\end{equation}
\end{prop}

Two kinds of reductions of the $q$-KPHSCSs (\ref{eqns:exdKP-LaxEqn}) are also studied in \cite{Lin:ext-qKP},
these reductions give many (1+1)-dimensional $q$-deformed soliton equation with self-consistent sources,
e.g., the $q$-deformed KdV equation with sources, the $q$-deformed Boussinesq equation with sources.

For convenience,we write out some operators here
\begin{eqnarray*}
B_1=\p_q+ u_0, \qquad 
B_2=\p^2_q + v_1\p_q+ v_0,\qquad 
B_3=\p^3_q + {s}_2\p^2_q+ {s}_1\p_q+ {s}_0,
\end{eqnarray*}
\begin{eqnarray*}
\phi_i\partial_q^{-1}\psi_i =
r_{i1}\partial_q^{-1}+r_{i2}\partial_q^{-2}+r_{i3}\partial_q^{-3}+\ldots,
\qquad i=1,\ldots,N,
\end{eqnarray*}
where
\begin{eqnarray*}
v_1=\theta(u_0)+u_0, \qquad 
v_0= (\p_q u_0)+\theta(u_{1})+u_0^2+u_{1},\\ 
v_{-1}=(\p_q u_{1})+\theta(u_{2})+u_0u_{1}+u_{1}\theta^{-1}(u_0)+u_{2},
\end{eqnarray*}
\begin{eqnarray*}
{s}_2=\theta(v_1)+ u_0, \qquad 
 {s}_1=(\p_q v_1) + \theta(v_0)+ u_0v_1+u_{1}, \\ 
 {s}_0=(\p_q v_0) + \theta(v_{-1}) + u_0v_0+u_{1}\theta^{-1}(v_{1})+u_{2}. 
\end{eqnarray*}
\begin{eqnarray*}
 {r}_{i1}=\phi_i\theta^{-1}(\psi_i), \qquad 
 {r}_{i2}=-\frac{1}{q} \phi_i \theta^{-2}(\partial_q\psi_i), \qquad 
 {r}_{i3}=\frac{1}{q^3}\phi_i\theta^{-3}(\partial_q^2\psi_i). 
\end{eqnarray*}
and $v_{-1}$ comes from $L^2=B_2+ v_{-1}\p^{-1}_q+
v_{-2}\p^{-2}_q+\cdots$.

Then, one can compute the following commutators
\begin{equation*}
     [B_2,B_3]=f_2\partial_q^2+f_1\partial_q+f_0, \qquad
        [B_2,\phi_i\partial_q^{-1}\psi_i]=g_{i1}\partial_q+g_{i0}+\ldots,
\end{equation*}
\begin{equation*}
    [B_3,\phi_i\partial_q^{-1}\psi_i]=h_{i2}\partial_q^2+h_{i1}\partial_q+h_{i0}+\ldots,\qquad i=1,\ldots,N,
\end{equation*} where
\begin{eqnarray*}
\fl
 f_2=\partial_q^2 s_2 
+(q+1)\theta(\partial_q s_1) +\theta^2(s_0) 
+v_1 \partial_q s_2 
+v_1 \theta(s_1) 
+v_0 s_2 
-(q^2+q+1) \theta(\partial_q^2 v_1) \\ 
-(q^2+q+1) \theta^2(\partial_q v_0 ) 
-(q+1) s_2 \theta(\partial_q v_1 ) 
-s_2 \theta^2(v_0) 
-s_1 \theta(v_1)-s_0, 
\end{eqnarray*}
\begin{eqnarray*}
\fl
 f_1=\partial_q^2 s_1 +(q+1)\theta(\partial_q s_0 ) 
+v_1\partial_q s_1 
+v_1\theta(s_0) 
+v_0s_1 -\partial_q^3 v_1 
-(q^2+q+1)\theta( \partial_q^2 v_0 ) \\
-s_2\partial_q^2 v_1  
-(q+1)s_2\theta( \partial_q v_0 ) 
-s_1\partial_q v_1  
-s_1\theta( v_0) 
-s_0v_1, 
\end{eqnarray*}
\begin{eqnarray*}
\fl
 f_0=\partial_q^2  s_0  +v_1 \partial_q s_0  
-\partial_q^3 v_0  -s_2 \partial_q^2 v_0  
-s_1 \partial_q v_0, 
\end{eqnarray*}
\begin{eqnarray*}
\fl
 g_{i1}=\theta^2(r_{i1}) -r_{i1},
\qquad
     g_{i0}=(q+1) \theta( \partial_q r_{i1} )
+\theta^2(r_{i2}) 
+v_1 \theta( r_{i1}) 
-r_{i1} \theta^{-1}(v_1) 
-r_{i2}, 
\end{eqnarray*}
\begin{eqnarray*}
\fl
 h_{i2}= \theta^3(r_{i1})-r_{i1},  
\qquad
     h_{i1}=(q^2+q+1) \theta^2(\partial_q r_{i1} )  
+\theta^3(r_{i2})+s_2 \theta^2(r_{i1}) 
-r_{i1} \theta^{-1}(s_2). 
\end{eqnarray*}
\begin{eqnarray*}
\fl
 h_{i0}=(q^2+q+1) \theta( \partial_q^2 r_{i1} ) 
+(q^2+q+1) \theta^2(\partial_q r_{i2} )
+\theta^3(r_{i3}) 
+(q+1) s_2  \theta( \partial_q r_{i1} )  \\ 
+s_2  \theta^2(r_{i2}) 
+s_1  \theta( r_{i1}) 
-r_{i1} \theta^{-1}( s_1 ) 
+\frac{1}{q} r_{i1} \theta^{-2}(\partial_q s_2 ) 
-r_{i2} \theta^{-2}(s_2 ) 
-r_{i3}. 
\end{eqnarray*}
Now, we list some examples in the $q$-KPHSCSs
(\ref{eqns:extdkp}) \cite{Lin:ext-qKP}.

\begin{eg}
  The first type of $q$-deformed KP equation with self-consistent sources (1{\it{st}}-$q$-KPSCS) is given by  (\ref{eqns:extdkp}) with  $n=2$ and $k=3$
\numparts
    \label{eqns:exam1}
    \begin{eqnarray}
      &-\frac{\partial s_2}{\partial t_2}+f_2=0, \label{eqns:exam1a}\\
      &\frac{\partial v_1}{\partial \tau_3}-\frac{\partial s_1}{\partial
      t_2}+f_1+\sum_{i=1}^{N}g_{i1}=0,\label{eqns:exam1b}\\
      &\frac{\partial v_0}{\partial \tau_3}-\frac{\partial s_0}{\partial
      t_2}+f_0+\sum_{i=1}^{N}g_{i0}=0,\label{eqns:exam1c}\\
      &\phi_{i,t_2}=B_2(\phi_{i}),\quad \psi_{i,t_2}=-B_2^{*}(\psi_{i}),\quad i=1,2,\cdots,N. \label{eqns:exam1d}
         \end{eqnarray}
\endnumparts

Let $q\rightarrow 1$ and $u_0\equiv 0$, then the 1{\it{st}}-$q$-KPSCS
 reduces to the first type of KP equation with
self-consistent sources \cite{Melnikov1983,Melnikov1987,Lin:ext-qKP}.
\end{eg}

\begin{eg}
  The second type of $q$-deformed KP equation with self-consistent sources (2{\it{nd}}-$q$-KPSCS) is given by  (\ref{eqns:extdkp}) with  $n=3$ and $k=2$
\numparts
    \label{eqns:exam2}
    \begin{eqnarray}
      &\frac{\partial s_2}{\partial \tau_2}-f_2+\sum_{i=1}^{N}h_{i2}=0,\label{eqns:exam2a}\\
      &\frac{\partial s_1}{\partial \tau_2}-\frac{\partial v_1}{\partial
      t_3}-f_1+\sum_{i=1}^{N}h_{i1}=0,\label{eqns:exam2b}\\
      &\frac{\partial s_0}{\partial \tau_2}-\frac{\partial v_0}{\partial
      t_3}-f_0+\sum_{i=1}^{N}h_{i0}=0,\label{eqns:exam2c}\\
      &\phi_{i,t_3}=B_3(\phi_{i}),\quad \psi_{i,t_3}=-B_3^{*}(\psi_{i}),\quad i=1,2,\cdots,N. \label{eqns:exam2d}
         \end{eqnarray}
\endnumparts
 Let $q\rightarrow 1$ and $u_0\equiv 0$, then the 2{\it{nd}}-$q$-KPSCS
 reduces to the second type of KP equation with
self-consistent sources \cite{Melnikov1983,Lin:ext-qKP}.
\end{eg}


\section{Generalized dressing approach for the $q$-KPHSCSs}

We will first give the dressing approach for the
$q$-KPH (\ref{eqn:dKP-LaxEqn}). Assume that the operator $L$ of $q$-KP
hierarchy (\ref{eqn:dKP-LaxEqn}) can be written as a dressing form
\begin{equation}
  \label{eqn:dress}
  L=S\pq S^{-1},
\end{equation}
with  $ S = \pq^{N} + w_1\pq^{N-1} + w_2\pq^{N-2}+\cdots + w_N. $

It is easy to
  verify 
   that if $S$ satisfies the Sato equation
   \begin{equation}
     \label{eqn:W-evolution-dKP}
     S_{t_n}=-L^{n}_-S,
   \end{equation}
then  $L$ defined by (\ref{eqn:dress}) satisfies the $q$-KP hierarchy (\ref{eqn:dKP-LaxEqn}).

If there are $N$ linearly independent functions $h_1,\ldots, h_N$ solving
$S(h_i)=0$, then $w_1,\ldots, w_N$ are completely determined by
solving the linear equations
    $$ \left( \begin{array}{cccc}
    h_1 & \pq(h_1) & \cdots & \pq^{N-1}(h_1)\\
    h_2 & \pq(h_2) & \cdots & \pq^{N-1}(h_2)\\
    \vdots & \vdots & \vdots & \vdots\\
    h_N & \pq(h_N) & \cdots & \pq^{N-1}(h_N)
  \end{array}
  \right)
 \left( \begin{array}{c}
    w_N \\
    w_{N-1}\\
    \vdots\\
    w_1
  \end{array}
  \right)
 =-
 \left( \begin{array}{c}
    \pq^N(h_1)\\
    \pq^N(h_2)\\
    \vdots\\
    \pq^N(h_N)
  \end{array}
  \right). $$
Then the operator $S$ can be written as
\begin{equation}
  \label{eqn:W}
  S = \frac{1
  }{{\rm{Wrd}}(h_1,\cdots,h_N)}\left|
    \begin{array}{ccccc}
      h_1 & h_2 & \cdots & h_N & 1\\
      \pq(h_1) & \pq(h_2) & \cdots & \pq(h_N) & \pq\\
      \vdots & \vdots & \vdots & \vdots & \vdots \\
      \pq^N(h_1) & \pq^N(h_2) & \cdots & \pq^N(h_N) & \pq^N
    \end{array}\right|,
\end{equation}
where
    $${\rm{Wrd}}(h_1,\cdots,h_N)=\left|
    \begin{array}{cccc}
      h_1 & h_2 & \cdots & h_N \\
       \pq(h_1) & \pq(h_2) & \cdots & \pq(h_N)\\
      \vdots & \vdots & \vdots & \vdots \\
       \pq^{N-1}(h_1) & \pq^{N-1}(h_2) & \cdots & \pq^{N-1}(h_N)
    \end{array}\right|. $$
\begin{rmk}
 The denominator of $S$ (\ref{eqn:W}) is actually a $q$-deformed
 Wronskian determinant, so we may denote it as $\Wr(h_1,\cdots,h_N)$.
 The numerator of $S$ (\ref{eqn:W}) is a formal determinant, which is
  denoted by $\Wr(h_1,\cdots,h_N,\pq)$. It is understood as an expansion with
  respect to its last column, in which all sub-determinants are collected on the
  left of the difference operator $\pq^j$.
\end{rmk}

Then, we have the dressing approach for the $q$-KP hierarchy (\ref{eqn:dKP-LaxEqn}) as the following.

\begin{prop}
\label{prop:2}
Assume that $h_i$ satisfies
\begin{equation}
\label{eq:hi}
  h_{i,t_n}=\pq^n(h_i),\qquad i=1,\cdots,N,
\end{equation}
and $S$ and $L$ are constructed as (\ref{eqn:W}) and
(\ref{eqn:dress}) respectively, then  $S$ and $L$ satisfy
the Sato equation (\ref{eqn:W-evolution-dKP}) and the $q$-KP hierarchy (\ref{eqn:dKP-LaxEqn}).
\end{prop}

\noindent
{\bf{Proof }}
   Apply partial derivative $\partial_{t_n}$ to the
  equation $S(h_i)=0$, and notice that $h_i$'s are linearly independent, then we have $S_{t_n}+L^n_-S=0$.
This completes the proof.

Unfortunately, the dressing approach given above can not provide the
evolution with respect to the new variable $\tau_k$. Now we
will generalize the dressing approach to solve the $q$-KPHSCSs (\ref{eqns:exdKP-LaxEqn}) and give
exact formulas for $\phi_i$ and $\psi_i$. First, we have the following
lemma.

\begin{lemm}
  \label{lm:2}
  For any q-pseudo-operator $S$, if $S$  satisfies
\numparts
  \label{eq:26}
  \begin{eqnarray}
   S_{t_n}&=-L^n_-S\label{eq:26a}\\
   S_{\ta_k}&=-L^k_-S+\sum_{i=1}^N\phi_i\pq^{-1}\psi_iS.\label{eq:26b}
  \end{eqnarray}
\endnumparts
then $L$ defined by (\ref{eqn:dress}) satisfies (\ref{eqn:exdKP-LaxEqna}) and
(\ref{eqn:exdKP-LaxEqnb}).
\end{lemm}
\noindent
{\bf{Proof }}
We only write out the proof for (\ref{eqn:exdKP-LaxEqna}).
 \begin{eqnarray*}
 \fl
    L_{\ta_k}=S_{\ta_k}\pq S^{-1}-S\pq S^{-1} S_{\ta_k} S^{-1}
             =(-L^k_-+\sum_i \phi_i\pq^{-1}\psi_i)L+ L(L^k_--\sum_i\phi_i\pq^{-1}\psi_i) \\
    =[-L^k_-+\sum_i \phi_i\pq^{-1}\psi_i,L]
             =[B_k+\sum_{i=1}^N \phi_i\pq^{-1}\psi_i,L].
  \end{eqnarray*}

 This dressing operator $S$ can be constructed as following. Let
$f_i$ and $g_i$ satisfy
\numparts
  \label{eq:fg}
  \begin{eqnarray}
    &f_{i,t_n}=\pq^n(f_i),\quad f_{i,\ta_k}=\pq^k(f_i)\quad \\
    &g_{i,t_n}=\pq^n(g_i),\quad g_{i,\ta_k}=\pq^k(g_i),\quad i=1,\ldots,N.
  \end{eqnarray}
\endnumparts
And let $h_i$ be the linear combination of $f_i$ and $g_i$ as
\begin{equation}
  \label{eq:h'}
  h_i=f_i+\af_i(\ta_k)g_i,\qquad i=1,\ldots,N,
\end{equation}
with the coefficient $\af_i$ being a differentiable function of
$\ta_k$. Suppose $h_1,\ldots, h_N$ are linearly independent. Then
clearly $S$ defined by (\ref{eqn:W}) and (\ref{eq:h'}) still satisfy
(\ref{eqn:W-evolution-dKP}) according to Proposition \ref{prop:2}.
To claim that $S$ satisfy (\ref{eq:26b}), we present
\begin{equation}
  \label{eqn:EigenFns}
  \phi_i=-\dot{\af}_i S(g_i),\qquad
  \psi_i=(-1)^{N-i}\tht\left(\frac{\Wr( h_1,\cdots,\hat{ h}_i,\cdots, h_N)}
  {\Wr( h_1,\cdots, h_N)}\right),
\end{equation}
where the hat $\hat{\;}$ means to rule out this term from the
$q$-deformed Wronskian determinant, and
$\dot{\af}_i=\frac{d\af_i}{d\ta_k}$. Then we
have the generalized dressing approach for the $q$-KPHSCSs (\ref{eqns:exdKP-LaxEqn}) as the following proposition.

\begin{prop}
  \label{prop:3}
  Let $S$ be defined by (\ref{eqn:W}) and (\ref{eq:h'}), $L=S\pq S^{-1}$, $\phi_{i}$ and $\psi_{i}$
  be given by (\ref{eqn:EigenFns}), then $S$ satisfies (\ref{eq:26}) and $L$,
  $\phi_i$, $\psi_i$ satisfy the $q$-KPHSCSs (\ref{eqns:exdKP-LaxEqn}).
\end{prop}

To prove Proposition \ref{prop:3}, we need several lemmas. The first
one is the $q$-deformed version of Oevel and Strampp's lemma  \cite{OS96}.

\begin{lemm}
  \label{lm:OS}
  (\it Oevel and Strampp \cite{OS96}) Under the condition of Proposition \ref{prop:3}, we have
  $$S^{-1}=\sum_{i=1}^N h_i\pq^{-1} \psi_i.$$
\end{lemm}

\begin{lemm}
  \label{lm:vanish}
  Under the condition of Proposition \ref{prop:3}, we have
   $\pq^{-1}S^*(\psi_i)=0,$ for $i=1,\ldots,N$.
\end{lemm}
\noindent
{\bf{Proof }}
   It can be proved that
   \begin{equation}
   \label{eq:34}
   (\pq^{-1}\psi_i S)_{-}=\pq^{-1}S^{*}(\psi_i).
   \end{equation}
  Using Lemma \ref{lm:OS}, we have
  \begin{eqnarray*}
  0&=(\pq^jS^{-1}\circ S)_{-}=(\pq^j\sum_{i=1}^Nh_i\pq^{-1}\psi_iS)_{-}
         =(\sum_{i=1}^N\pq^j(h_{i})\pq^{-1}\psi_iS)_{-}\\
   &=\sum_{i=1}^N\pq^j(h_i)\pq^{-1} S^*(\psi_i),\quad j=0,1,2,\ldots.
  \end{eqnarray*}
 Solving the  equations with respect to
  $\pq^{-1}S^*(\psi_i)$, we get Lemma \ref{lm:vanish}.

\begin{lemm}
  \label{lm:key}
  Under the condition of Proposition \ref{prop:3},
  the operator $\pq^{-1} \psi_i S$ is a non-negative difference operator
  and
  \begin{equation}
    \label{eq:key-eq}
    (\pq^{-1} \psi_i S)(h_j)=\dt_{ij},\quad 1\le i,j\le N.
  \end{equation}
\end{lemm}
\noindent
{\bf{Proof }}
 Lemma \ref{lm:vanish} and (\ref{eq:34}) implies that
$\pq^{-1}\phi_{i}S$ is a non-negative difference operator.\\
We define functions
  $c_{ij}=(\pq^{-1}\psi_iS)(h_j)$,~
 then  $\pq(c_{ij})=\psi_iS(h_j)=0$, which means $c_{ij}$ does not depend on
 $x$ in the sense of $q$-deformed version. From Lemma \ref{lm:OS}, we find that
 \begin{eqnarray*}
        \sum_{i=1}^N \pq^k(h_i)c_{ij}
    & =\sum_{i=1}^N \pq^k(h_ic_{ij})=\pq^k(\sum_{i=1}^N h_ic_{ij})
        =\pq^k(\sum_i^Nh_i\pq^{-1} \psi_i S(h_j))\\
   & =\pq^k(S^{-1}S)(h_j) =\pq^k(h_j),
 \end{eqnarray*}
  since the functions $h_1,h_2,\cdots,h_N$ are linearly independent,
  we can easily conclude that $c_{ij}=\dt_{ij}$.\\

\noindent {\sl Proof of Proposition \ref{prop:3}.}\\
  \indent Analogous to the proof of Proposition \ref{prop:2}, we can prove (\ref{eq:26a}).
  For (\ref{eq:26b}), taking $\partial_{\ta_k}$ to the identity
  $S(h_i)=0$,we find
  \begin{eqnarray*}
    0&=(S_{\ta_k})(h_i)+(S\pq^k)(h_i)+\dot\af_i S(g_i)
    =(S_{\ta_k})(h_i)+(L^kS)(h_i)-\sum_{j=1}^N\phi_j\dt_{ji}\\
    &=(S_{\ta_k}+L^k_-S-\sum_{j=1}^N\phi_j\pq^{-1}\psi_jS)(h_i).
  \end{eqnarray*}
  Obviously, $S_{\tau_k}+L^k_-S$ is a pure
  difference operator of degree $<N$. and moreover using Lemma \ref{lm:key},
  $\sum\limits_{j=1}^N\phi_j\pq^{-1}\psi_jS$ is also a pure
  difference operator of degree $<N$. so $S_{\ta_k}+L^k_-S-\sum\limits_{j=1}^N\phi_j\pq^{-1}\psi_jS$
  is a pure difference operator of degree $<N$.
  Since the non-negative difference operator acting on $h_i$ in the last
  expression has degree $<N$, it can not annihilate $N$ independent functions
  unless the operator itself vanishes.

  Hence (\ref{eq:26}) is
  proved. Then Lemma \ref{lm:2} leads to (\ref{eqn:exdKP-LaxEqna})
  and
  (\ref{eqn:exdKP-LaxEqnb}).

   The proof of the first equation in (\ref{eqn:exdKP-LaxEqnc}) is
  the following.
  \begin{eqnarray*}
  \phi_{i,t_n}&=-\dot{\af_i}(S(g_i))_{t_n}=-\dot{\af}(S_{t_n}+S\pq^n)(g_i)\\
             &=-\dot{\af_i}(-L^n_-S+L^nS)(g_i)=-\dot{\af}B_nS(g_i)=B_n(\phi_i).
  \end{eqnarray*}

  \noindent And it remains
  to prove the second equation in (\ref{eqn:exdKP-LaxEqnc}). Firstly, we see that
  \begin{eqnarray*}
  (S^{-1})_{t_n}&=((S^{-1})_{t_n})_-=(-S^{-1}S_{t_n}S^{-1})_-=(S^{-1}(L^n-B_n))_-\\
                &=(\pq^nS^{-1})_--(S^{-1}B_n)_-=(\pq^n\sum^N_{i=1}h_i\pq^{-1}\psi_i)_--(\sum^N_{i=1}h_i\pq^{-1}\psi_iB_n)_-\\
                &=\sum^N_{i=1}\pq^n(h_i)\pq^{-1}\psi_i-\sum^N_{i=1}h_i\pq^{-1}B_n^*(\psi_i).
  \end{eqnarray*}
  On the other hand,
  $(S^{-1})_{t_n}=(\sum\limits^N_{i=1}h_i\pq^{-1}\psi_i)_{t_n}=\sum\limits^N_{i=1}\pq^n(h_i)\pq^{-1}\psi_i-\sum\limits^N_{i=1}h_i\pq^{-1}\psi_{i,t_n}$,
  so we have $\sum\limits^N_{i=1}h_i\pq^{-1}(B_n^*(\psi_i)+\psi_{i,t_n})=0$.
Since $h_i$, $i=1,...,N$ are linearly independent, it is easy to get $\psi_{i,t_n}=-B_n^*(\psi_i)$.

Thus, we proved Proposition \ref{prop:3} (the generalized dressing approach for the $q$-KPHSCSs (\ref{eqns:exdKP-LaxEqn})).


\section{The $q$-mKP hierarchy with self-consistent sources ($q$-mKPHSCSs)}
\label{sec:qmKP} In this section, we will construct the $q$-deformed mKP hierarchy with self-consistent sources ($q$-mKPHSCSs).
The Lax operator $\widetilde{L}$ of $q$-mKP hierarchy is
defined by
$$\widetilde{L}=\widetilde{u}\pq+\widetilde{u}_0+\widetilde{u}_1\pq^{-1}+\widetilde{u}_2\pq^{-2}+\cdots.$$
And the Lax equation of $q$-mKP hierarchy is given by
\begin{equation}
  \label{meqn:1}                                         
  \widetilde{L}_{t_n}=[\widetilde{B}_n,\widetilde{L}],\qquad \widetilde{B}_n=(\widetilde{L}^n)_{\geq 1}.
\end{equation}
The $\p_{t_n}$-flows are commutative with each other, and we can
easily deduce the zero-curvature equation
 \begin{equation}
 \label{meqn:2}                                          
  \widetilde{B}_{n,t_m}-\widetilde{B}_{m,t_n}+[\widetilde{B}_n,\widetilde{B}_m]=0.
\end{equation}
When $n=2$ and $m=3$, we get the $q$-mKP equation. If we take
$q\rightarrow 1$ and $\widetilde{u}\equiv 1$, then the $q$-mKP equation will
reduce to the mKP equation
    $$4v_t-v_{xxx}+6v^2v_x-3(D^{-1}v_{yy})-6v_x(D^{-1}v_y)=0,$$
where $t:=t_3$, $y:=t_2$, $v:=\widetilde{u}_0$.

According to the squared eigenfunction symmetry (see \cite{OS93,OC98} and the references therein), we can construct
a $q$-mKP hierarchy with self-consistent sources ($q$-mKPHSCSs) as
\numparts
  \label{meqn:3}                                        
  \begin{eqnarray}
    \widetilde{L}_{\ta_k}&=[\widetilde{B}_k+\sum_{i=1}^N\widetilde{\phi}_i\pq^{-1}\widetilde{\psi}_i\pq,\widetilde{L}], \label{meqn:3a}\\
    \widetilde{L}_{t_n}&=[\widetilde{B}_n,\widetilde{L}], \quad \forall n\neq k, \label{meqn:3b}\\
    \widetilde{\phi}_{i,t_n}&=\widetilde{B}_n(\widetilde{\phi}_i), \label{meqn:3c}\\
    \widetilde{\psi}_{i,t_n}&=-(\pq \widetilde{B}_n \pq^{-1})^*(\widetilde{\psi}_i),~i=1,\cdots,N. \label{meqn:3d}
  \end{eqnarray}
\endnumparts
Then it is easy to get the
zero curvature equation for the $q$-mKPHSCSs (\ref{meqn:3})
\begin{eqnarray}
\label{meqn:4}                                           
  \widetilde{B}_{n,\tau_{k}}-(\widetilde{B}_{k}+\sum_{i=1}^{N}\widetilde{\phi}_{i}\pq^{-1}\widetilde{\psi}_{i}\pq)_{t_{n}}+[\widetilde{B}_{n},\widetilde{B}_{k}
      +\sum_{i=1}^{N}\widetilde{\phi}_{i}\pq^{-1}\widetilde{\psi}_{i}\pq]=0.
\end{eqnarray}
Under the condition (\ref{meqn:3c}) and (\ref{meqn:3d}), the Lax
pair for the $q$-mKPHSCSs (\ref{meqn:3}) is given by
$$\Psi_{t_n}=\widetilde{B}_n(\Psi),\qquad \Psi_{\tau_k}=(\widetilde{B}_k+\sum\limits^N_{i=1}\widetilde{\phi}_i\pq^{-1}\widetilde{\psi}_i\pq)(\Psi).$$

First, for convenience, we write out some operators here
\begin{eqnarray*}
&\widetilde{B}_1=\widetilde{u}\p_q, \qquad \widetilde{B}_2=\widetilde{v}_2\p^2_q + \widetilde{v}_1\p_q,\qquad  \widetilde{B}_3=\widetilde{s}_3\p^3_q +{\widetilde{s}}_2\p^2_q+ {\widetilde{s}}_1\p_q,\\
&\widetilde{\phi}_i\partial_q^{-1}\widetilde{\psi}_i\p_q=\widetilde{r}_{i0}+\widetilde{r}_{i1}\partial_q^{-1}+\widetilde{r}_{i2}\partial_q^{-2}+\cdots,\qquad i=1,\ldots,N,
\end{eqnarray*}

\noindent where
\begin{eqnarray*}
&\widetilde{v}_2=\widetilde{u}\theta(\widetilde{u}),\qquad
\widetilde{v}_1=\widetilde{u}(\theta(\widetilde{u}_0)+\widetilde{u}_0+\pq(\widetilde{u})), \\
&\widetilde{v}_0= \widetilde{u}_1\theta^{-1}(\widetilde{u})+\widetilde{u}_0^2+\widetilde{u}\theta(\widetilde{u}_{1})+\widetilde{u}\pq(\widetilde{u}_0),\\
&{\widetilde{s}}_3=\widetilde{u}\theta(\widetilde{v}_2), \qquad
{\widetilde{s}}_2=\widetilde u \partial_q (\widetilde v_2)+\widetilde u \theta(\widetilde v_1)+\widetilde u_0 \widetilde v_2,\\ 
&{\widetilde{s}}_1=\widetilde u\partial_q(\widetilde v_1)+\widetilde u \theta(\widetilde v_0)+\widetilde u_0 \widetilde v_1+\widetilde u_1 \theta^{-1}(\widetilde v_2), \\ 
&{\widetilde{r}}_{i0}=\widetilde{\phi}_i\theta^{-1}(\widetilde{\psi}_i),\qquad
{\widetilde{r}}_{i1}=-\frac{1}{q} \widetilde{\phi}_i\theta^{-2}(\partial_q\widetilde{\psi}_i),\qquad
{\widetilde{r}}_{i2}=\frac{1}{q^3}\widetilde{\phi}_i\theta^{-3}(\partial_q^2\widetilde{\psi}_i),
\end{eqnarray*}
and $\widetilde{v}_{0}$ comes from $\widetilde{L}^2=\widetilde{B}_2+
\widetilde{v}_{0}+\widetilde{v}_{-1}\p^{-1}_q+\cdots$.

Then, one can compute the following commutators
\begin{eqnarray*}
     &[\widetilde{B}_2,\widetilde{B}_3]=\widetilde{f}_3\partial_q^3+\widetilde{f}_2\partial_q^2+\widetilde{f}_1\pq, \qquad
     [\widetilde{B}_2,\widetilde{\phi}_i\partial_q^{-1}\widetilde{\psi}_i\pq]=\widetilde{g}_{i2}\partial_q^2+\widetilde{g}_{i1}\pq+\cdots,\\
     &[\widetilde{B}_3,\widetilde{\phi}_i\partial_q^{-1}\widetilde{\psi}_i\pq]=\widetilde{h}_{i3}\partial_q^3+\widetilde{h}_{i2}\partial_q^2+\widetilde{h}_{i1}\pq+\cdots,\qquad i=1,\ldots,N.
\end{eqnarray*}
where
\begin{eqnarray*}
\fl
\widetilde{f}_3=\widetilde{v}_2\pq^2(\widetilde{s}_3)+(q+1)\widetilde{v}_2\theta(\pq(\widetilde{s}_2))
               +\widetilde{v}_2\theta^2(\widetilde{s}_1)+\widetilde{v}_1\pq(\widetilde{s}_3)+\widetilde{v}_1\tht(\widetilde{s}_2)
               -(q^2+q+1)\widetilde{s}_3\tht(\pq^2(\widetilde{v}_2))\\
         -(q^2+q+1)\widetilde{s}_3\tht^2(\pq(\widetilde{v}_1))-(q+1)\widetilde{s}_2\tht(\pq(\widetilde{v}_2))
               -\widetilde{s}_2\tht^2(\widetilde{v}_1)-\widetilde{s}_1\tht(\widetilde{v}_2),\\
\fl
\widetilde{f}_2=\widetilde{v}_2\pq^2(\widetilde{s}_2)+(q+1)\widetilde{v}_2\theta(\pq(\widetilde{s}_1))
              +\widetilde{v}_1\pq(\widetilde{s}_2)+\widetilde{v}_1\theta(\widetilde{s}_1)-\widetilde{s}_3\pq^3(\widetilde{v}_2)\\
              -(q^2+q+1)\widetilde{s}_3\theta( \pq^2 (\widetilde{v}_1))
            -\widetilde{s}_2\pq^2(\widetilde{v}_2)-(q+1)\widetilde{s}_2\theta(\pq (\widetilde{v}_1))
              -\widetilde{s}_1\pq(\widetilde{v}_2) -\widetilde{s}_1\theta( \widetilde{v}_1), \\
\fl
\widetilde{f}_1=\widetilde{v}_2\pq^2(\widetilde{s}_1)+
            \widetilde{v}_1\pq(\widetilde{s}_1)-\widetilde{s}_3\pq^3(\widetilde{v}_1)
              -\widetilde{s}_2\pq^2(\widetilde{v}_1)-\widetilde{s}_1 \pq(\widetilde{v}_1),\\
\fl
\widetilde{g}_{i2}=\widetilde{v}_2[\theta^2(\widetilde{r}_{i0})-\widetilde{r}_{i0}], \\
\fl
\widetilde{g}_{i1}=(q+1)\widetilde{v}_2\theta(\pq(\widetilde{r}_{i0}))+\widetilde{v}_2\theta^2(\widetilde{r}_{i1})
              +\widetilde{v}_1 \theta( \widetilde{r}_{i0}) -\widetilde{r}_{i0}\widetilde{v}_1-\widetilde{r}_{i1}\theta^{-1}(\widetilde{v}_2), \\
\fl
\widetilde{h}_{i3}=\widetilde{s}_3[\theta^3(\widetilde{r}_{i0})-\widetilde{r}_{i0}],\\
\fl
\widetilde{h}_{i2}=(q^2+q+1)\widetilde{s}_3\theta^2(\partial_q
    (\widetilde{r}_{i0}))+\widetilde{s}_3\theta^3(\widetilde{r}_{i1})
              +\widetilde{s}_2 \theta^2(\widetilde{r}_{i0})-\widetilde{r}_{i0}\widetilde{s}_2 -\widetilde{r}_{i1}\theta^{-1}(\widetilde{s}_3),\\
\fl
\widetilde{h}_{i1}=(q^2+q+1)\widetilde{s}_3\theta(\partial_q^2(\widetilde{r}_{i0}))+(q^2+q+1)\widetilde{s}_3\theta^2(\partial_q(\widetilde{r}_{i1}))
              +\widetilde{s}_3\theta^3(\widetilde{r}_{i2})+(q+1)\widetilde{s}_2\theta(\partial_q(\widetilde{r}_{i0}))  \\
        +\widetilde{s}_2\theta^2(\widetilde{r}_{i1})+\widetilde{s}_1\theta(\widetilde{r}_{i0})
              -\widetilde{r}_{i0}\widetilde{s}_1 +\frac{1}{q} \widetilde{r}_{i1} \theta^{-2}(\pq(\widetilde{s}_3))-\widetilde{r}_{i1}\tht^{-1}(\widetilde{s}_2)
      -\widetilde{r}_{i2}\theta^{-2}(\widetilde{s}_3).
\end{eqnarray*}
Now, we can list the two types of $q$-mKP equations with self-consistent source.

\begin{eg}
When $n=2$ and $k=3$, the $q$-mKPHSCSs (\ref{meqn:3}) gives the first type of $q$-mKP equation with self-consistent sources (1\it{st}-$q$-mKPSCS)
\numparts
    \label{meqn:exam1}                                                         
    \begin{eqnarray}
      &-\frac{\partial \widetilde{s}_3}{\partial t_2}+\widetilde{f}_3=0, \label{meqns:exam1a}\\
      &\frac{\partial \widetilde{v}_2}{\partial \tau_3}-\frac{\partial \widetilde{s}_2}{\partial
      t_2}+\widetilde{f}_2+\sum_{i=1}^{N}\widetilde{g}_{i2}=0,\label{meqns:exam1b}\\
      &\frac{\partial \widetilde{v}_1}{\partial \tau_3}-\frac{\partial \widetilde{s}_1}{\partial
      t_2}+\widetilde{f}_1+\sum_{i=1}^{N}\widetilde{g}_{i1}=0,\label{meqns:exam1c}\\
      &\widetilde{\phi}_{i,t_2}=\widetilde{B}_2(\widetilde{\phi}_{i}),\quad
      \widetilde{\psi}_{i,t_2}=-(\pq \widetilde{B}_2\pq^{-1})^{*}(\widetilde{\psi}_{i}),\quad i=1,2,\ldots,N. \label{meqns:exam1d}
         \end{eqnarray}
\endnumparts

Let $q\rightarrow 1$ and $u\equiv 1$, then the first type of $q$-mKP equation with self-consistent source (\ref{meqn:exam1}) reduces to the first type of mKP equation with
self-consistent sources which reads as
\numparts
    \begin{eqnarray*}
     &4\widetilde{u}_{0,t}-\widetilde{u}_{0,xxx}+6\widetilde{u}_0^2\widetilde{u}_{0,x}-3D^{-1}\widetilde{u}_{0,yy}-6\widetilde{u}_{0,x}D^{-1}\widetilde{u}_{0,y}+4\sum^N_{i=1}(\widetilde{\phi}_i\widetilde{\psi}_i)_x=0,\\
     &\widetilde{\phi}_{i,y}=\widetilde{\phi}_{i,xx}+2\widetilde{u}_0\widetilde{\phi}_{i,x}, \\
     &\widetilde{\psi}_{i,y}=-\widetilde{\psi}_{i,xx}+2\widetilde{u}_0\widetilde{\psi}_{i,x}, \qquad
     i=1,\ldots,N,
    \end{eqnarray*}
\endnumparts
where $t:=\tau_3, y:=t_2$. 
\end{eg}

\begin{eg}
 When $n=3$ and $k=2$, the $q$-mKPHSCSs (\ref{meqn:3}) gives
  the second type of $q$-mKP equation with self-consistent source (2\it{nd}-$q$-mKPSCS)
\numparts
    \label{meqns:exam2}                                               
    \begin{eqnarray}
      &\frac{\partial \widetilde{s}_3}{\partial \tau_2}-\widetilde{f}_3+\sum_{i=1}^{N}\widetilde{h}_{i3}=0,\label{meqns:exam2a}\\
      &\frac{\partial \widetilde{s}_2}{\partial \tau_2}-\frac{\partial \widetilde{v}_2}{\partial
      t_3}-\widetilde{f}_2+\sum_{i=1}^{N}\widetilde{h}_{i2}=0,\label{meqns:exam2b}\\
      &\frac{\partial \widetilde{s}_1}{\partial \tau_2}-\frac{\partial \widetilde{v}_1}{\partial
      t_3}-\widetilde{f}_1+\sum_{i=1}^{N}\widetilde{h}_{i1}=0,\label{meqns:exam2c}\\
      &\widetilde{\phi}_{i,t_3}=\widetilde{B}_3(\widetilde{\phi}_{i}),
      \quad \widetilde{\psi}_{i,t_3}=-(\pq \widetilde{B}_3\pq^{-1})^{*}(\widetilde{\psi}_{i}),\qquad i=1,\cdots,N. \label{meqns:exam2d}
         \end{eqnarray}
\endnumparts

 Let $q\rightarrow 1$ and $u\equiv 1$, then the second type of $q$-mKP equation with self-consistent source  (\ref{meqns:exam2}) reduces to the second type of mKP equation with self-consistent sources which reads as
\numparts
    \begin{eqnarray*}
\fl
     4\widetilde{u}_{0,t}-\widetilde{u}_{0,xxx}+6\widetilde{u}_0^2\widetilde{u}_{0,x}-3D^{-1}\widetilde{u}_{0,yy}-6\widetilde{u}_{0,x}D^{-1}\widetilde{u}_{0,y}\\
     +\sum^N_{i=1}[3(\widetilde{\phi}_i\widetilde{\psi}_{i,xx}-\widetilde{\phi}_{i,xx}\widetilde{\psi}_i)-3(\widetilde{\phi}_i\widetilde{\psi}_i)_y-6(\widetilde{u}_0\widetilde{\phi}_i\widetilde{\psi}_i)_x]=0,\\
\fl
     \widetilde{\phi}_{i,t}=\widetilde{\phi}_{i,xxx}+3\widetilde{u}_0\widetilde{\phi}_{i,xx}+\frac32(D^{-1}\widetilde{u}_{0,y})\widetilde{\phi}_{i,x}+\frac32\widetilde{u}_{0,x}\widetilde{\phi}_{i,x}+\frac32\widetilde{u}_0^2\widetilde{\phi}_{i,x}
                              +\frac32\sum^N_{j=1}(\widetilde{\phi}_j\widetilde{\psi}_j)\widetilde{\phi}_{i,x},\\
\fl
     \widetilde{\psi}_{i,t}=\widetilde{\psi}_{i,xxx}-3\widetilde{u}_0\widetilde{\psi}_{i,xx}+\frac32(D^{-1}\widetilde{u}_{0,y})\widetilde{\psi}_{i,x}-\frac32\widetilde{u}_{0,x}\widetilde{\psi}_{i,x}+\frac32\widetilde{u}_0^2\widetilde{\psi}_{i,x}
                              +\frac32\sum^N_{j=1}(\widetilde{\phi}_j\widetilde{\psi}_j)\widetilde{\psi}_{i,x},
    \end{eqnarray*}
\endnumparts
where $y:=\tau_2, t:=t_3.$ 
\end{eg}

\section{The gauge transformation between the $q$-KPHSCSs and the $q$-mKPHSCSs}
In this section, we will give a
 gauge transformation between the $q$-KPHSCSs and the $q$-mKPHSCSs.
\begin{prop}
\label{prop:4} Suppose $L$, $\phi_i$'s, and $\psi_i$'s satisfy the
$q$-KPHSCSs (\ref{eqns:exdKP-LaxEqn}),
and $f$ is a particular eigenfunction for the Lax pair
(\ref{eqn:exdKP-LaxPair}) of the $q$-KPHSCSs, i.e.,
    $$ f_{t_n}=B_n(f),\qquad
    f_{\tau_k}=(B_k+\sum\limits^N_{i=1}\phi_i\pq^{-1}\psi_i)(f),$$
then
\begin{equation}
\label{gauge}
    \widetilde{L}:= f^{-1}Lf,\quad \widetilde{\phi_{i}}:=f^{-1}\phi_i,\quad \widetilde{\psi_{i}}:=-\theta\pq^{-1}(f\psi_i)=(\pq^{-1})^*(f\psi_i),
\end{equation}
satisfy the $q$-mKPHSCSs (\ref{meqn:3}).
\end{prop}

\noindent
{\bf{Proof }}
Since $f$ is the eigenfunction of the Lax pair
(\ref{eqn:exdKP-LaxPair}) for the $q$-KPHSCSs,
then
\begin{eqnarray*}
\fl
\widetilde{L}_{t_n}=(f^{-1}Lf)_{t_n}=-f^{-1}B_n(f)f^{-1}Lf+f^{-1}[B_n,L]f+f^{-1}LB_n(f)\\
\fl
               =-f^{-1}B_n(f)\widetilde{L}+[f^{-1}B_nf,\widetilde{L}]+\widetilde{L}f^{-1}B_n(f)=[f^{-1}B_nf-f^{-1}B_n(f),\widetilde{L}]
               =[\widetilde{B}_n,\widetilde{L}],
\end{eqnarray*}
here it is used that
$\Delta:=f^{-1}B_nf-f^{-1}B_n(f)=f^{-1}[(L^nf)_{\geq 0}-(L^n)_{\geq
0}(f)]=f^{-1}((L^nf)_{\geq 1})
               =(f^{-1}L^nf)_{\geq 1}=\widetilde{L}^n_{\geq 1}$, and we denote $\widetilde{L}^n_{\geq 1}$ by $\widetilde{B}_n$.
Moreover, we have
\begin{eqnarray*}
\fl
\widetilde{L}_{\tau_k}=(f^{-1}Lf)_{\tau_k}
                  =-f^{-1}(B_k+\sum\limits^N_{i=1}\phi_i\pq^{-1}\psi_i)(f)f^{-1}Lf+f^{-1}[B_k+\sum\limits^N_{i=1}\phi_i\pq^{-1}\psi_i,L]f\\
                  \qquad  +f^{-1}L(B_k+\sum\limits^N_{i=1}\phi_i\pq^{-1}\psi_i)(f)\\
                  =[f^{-1}(B_k+\sum\limits^N_{i=1}\phi_i\pq^{-1}\psi_i)f-f^{-1}(B_k+\sum\limits^N_{i=1}\phi_i\pq^{-1}\psi_i)(f),\widetilde{L}]\\
                  =[\widetilde{B}_k,\widetilde{L}]+\sum^N_{i=1}[f^{-1}\phi_i\pq^{-1}\psi_if-f^{-1}\phi_i\pq^{-1}\psi_i(f),\widetilde{L}]\\
                  =[\widetilde{B}_k,\widetilde{L}]+\sum^N_{i=1}[\widetilde{\phi}_i\pq^{-1}\circ\pq^*(\widetilde{\psi}_i)
                               +\widetilde{\phi}_i\pq^{-1}\circ\pq\circ\theta^{-1}(\widetilde{\psi}_i),\widetilde{L}]\\
                  =[\widetilde{B}_k,\widetilde{L}]+\sum^N_{i=1}[\widetilde{\phi}_i\pq^{-1}\circ\widetilde{\psi}_i\pq,\widetilde{L}]
                                  =[\widetilde{B}_k+\sum^N_{i=1}\widetilde{\phi}_i\pq^{-1}\widetilde{\psi}_i\pq,\widetilde{L}],
\end{eqnarray*}
\begin{eqnarray*}
\fl
\widetilde{\phi}_{i,t_n}=-f^{-1}B_n(f)f^{-1}\phi_i+f^{-1}B_n(\phi_i)=-f^{-1}B_n(f)\widetilde{\phi}_i+f^{-1}B_n(f\widetilde{\phi}_i)\\
                    =f^{-1}L^n_{\geq 0}f(\widetilde{\phi}_i)-f^{-1}L^n_{\geq 0}(f)\widetilde{\phi}_i
                            =(f^{-1}L^nf)_{\geq 0}(\widetilde{\phi}_i)-f^{-1}L^n_{\geq 0}(f)\widetilde{\phi}_i\\
                    =(f^{-1}L^nf)_{\geq 1}(\widetilde{\phi}_i)+(f^{-1}L^n_{\geq 0}(f))(\widetilde{\phi}_i)-f^{-1}L^n_{\geq 0}(f)\widetilde{\phi}_i
                            =\widetilde{B}_n(\widetilde{\phi}_i),
\end{eqnarray*}
\begin{eqnarray*}
\fl
\widetilde{\psi}_{i,t_n}=(\pq^{-1})^*[B_n(f)\psi_i-fB_n^*(\psi_i)]
                           =(\pq^{-1})^*[B_n(f)f^{-1}\pq^*(\widetilde{\psi}_i)-fB_n^*f^{-1}\pq^*(\widetilde{\psi}_i)]\\
                    =-(\pq^{-1})^*[((f(L^n)^*f^{-1})_{\geq 0}\pq^*(\widetilde{\psi}_i)-(L^n)_{\geq 0}(f)f^{-1}\pq^*(\widetilde{\psi}_i)]\\
                    =-(\pq^{-1})^*[((f^{-1}L^nf)_{\geq 0})^*\pq^*(\widetilde{\psi}_i)-(L^n)_{\geq 0}(f)f^{-1}\pq^*(\widetilde{\psi}_i)]\\
                    =-(\pq^{-1})^*((f^{-1}L^nf)_{\geq 1})^*\pq^*(\widetilde{\psi}_i)=-(\pq^{-1})^*\widetilde{B}_n^*\pq^*(\widetilde{\psi}_i)
                    =-(\pq\widetilde{B}_n\pq^{-1})^*(\widetilde{\psi}_i).\\
\end{eqnarray*}
This completes the proof.

Therefore, if a special eigenfunction $f$ for the Lax pair (\ref{eqn:exdKP-LaxPair}) of the
$q$-KPHSCSs is given, then we can get a solution of the
$q$-mKPHSCSs by the gauge transformation (\ref{gauge}). Here we choose
\begin{equation}
\label{meqn:5}                                              
    f=S(1)=(-1)^N\frac{\Wr(\pq(h_1),\pq(h_2),\cdots,\pq(h_N))}{\Wr(h_1,h_2,\cdots,h_N)}
\end{equation}
as the particular eigenfunction for the Lax pair (\ref{eqn:exdKP-LaxPair}) of the $q$-KPHSCSs, where $S$ is the dressing operator defined by
(\ref{eqn:W}) and (\ref{eq:h'}). Then the Wronskian solution for the
$q$-mKPHSCSs is
\numparts
\label{meqn:6}                                                       
  \begin{eqnarray}
 \fl
  \widetilde{L}=f^{-1}Lf=\frac{\Wr(h_1,
            \cdots,h_N,\pq)}{\Wr(\pq(h_1),
            \cdots,\pq(h_N))}
                               \pq\left[\frac{\Wr(h_1,
                               \cdots,h_N,\pq)}{\Wr(\pq(h_1),
                               \cdots,\pq(h_N))}\right]^{-1},\\
 \fl
  \widetilde{\phi_{i}}=f^{-1}\phi_i=-\dot{\alpha}_i\frac{\Wr(h_1,h_2,\cdots,h_N,g_i)}{\Wr(\pq(h_1),\pq(h_2),\cdots,\pq(h_N))},\\
 \fl
  \widetilde{\psi_{i}}=-\theta\pq^{-1}(f\psi_i)
                              =\theta\left(\frac{\Wr(\pq(h_1),\cdots,\hat{\pq(h_i)},\cdots,\pq(h_N))}{\Wr(h_1,h_2,\cdots,h_N)}\right),
                              \qquad i=1,\ldots, N.\label{meqn:6c}
  \end{eqnarray}
\endnumparts
The above expressions for $\widetilde{L}$ and $\widetilde{\phi}_i$'s can be
easily known by straightforward calculation, and the above
expressions for $\widetilde{\psi}_i$'s can be derived as follows.
First, we see
\begin{eqnarray*}
\fl
\sum^N_{i=1}\theta(h_i)\widetilde{\psi}_i=\sum^N_{i=1}\theta(-h_i\pq^{-1}(\psi_if))=\theta((\sum^N_{i=1}-h_i\pq^{-1}\psi_i)(f))
                                     =\theta(S^{-1}S(1))=1.
\end{eqnarray*}
And moreover we have the following relation (for $k\geq 1$),
\begin{eqnarray*}
\sum^N_{i=1}\theta(\pq^{k}(h_i))\widetilde{\psi}_i=\sum^N_{i=1}\theta[-\pq^{k}(h_i)\cdot\pq^{-1}(\psi_if)]\\
              =\sum^N_{i=1}\theta[-\pq(\pq^{k-1}(h_i)\cdot\pq^{-1}(\psi_if))+\theta(\pq^{k-1}(h_i)\cdot\psi_if)]\\
=\cdots
=\sum^N_{i=1}\theta[-\pq^{k}(h_i\pq^{-1}\psi_i(f))+\sum^{k-1}_{j=0}\pq^{k-j-1}(\theta(\pq^j(h_i))\psi_if)]\\
=\theta[\sum^{k-1}_{j=0}\pq^{k-j-1}(\sum^N_{i=0}\theta(\pq^j(h_i))\psi_if)].
\end{eqnarray*}
Notice the definition of $\psi_i$'s (\ref{eqn:EigenFns}) and
    $\sum\limits^N_{i=0}\theta(\pq^j(h_i))\psi_i=\delta_{j,N-1},$ $j=0,1,\ldots,N-1,$
then we have $\sum\limits^N_{i=1}\theta(\pq^k(h_i))\widetilde{\psi}_i=0,$ for
$k=1,\ldots,N-1$. Then using the Cramer principle, we can get the exact form of
$\widetilde{\psi}_i$'s (\ref{meqn:6}).

\section{Solutions of the $q$-KPHSCSs and the $q$-mKPHSCSs}
The generalized dressing approach (Proposition \ref{prop:3}) and the gauge
transformation (\ref{gauge}) give us a simple way to construct explicit solutions
of the $q$-KPHSCSs and the $q$-mKPHSCSs. Here we use the
first type of $q$-KP equation with self-consistent sources (\ref{eqns:exam1}) and the
first type of $q$-mKP equation with self-consistent sources (\ref{meqn:exam1}) as the
examples.

If we choose
\begin{eqnarray*}
 &f_i:=e_q(\lambda_ix)\exp(\lambda_i^2t_2+\lambda_i^3\ta_3)\equiv e_q(\lambda_ix)e^{\xi_i},\\
 &g_i:=e_q(\mu_ix)\exp(\mu_i^2t_2+\mu_i^3\ta_3)\equiv e_q(\mu_ix)e^{\et_i},\\
 &h_i:=f_i+\af_i(\ta_3)g_i=e_q(\lambda_ix)e^{\xi_i}+\af_i(\ta_3)e_q(\mu_ix)e^{\et_i}, \qquad i=1,\ldots,N,
\end{eqnarray*}
then the generalized dressing approach (Proposition \ref{prop:3})
enables us to get the soliton solutions to the first type of $q$-KP equation with sources (\ref{eqns:exam1}).

\begin{eg}
\label{solution-qKPSCS-I}
(One-soliton solution to the 1{\it{st}}-$q$-KPSCS (\ref{eqns:exam1})) Let
$N=1$, then
\begin{displaymath}
  S=\pq+w_0,\quad w_0=-\frac{\pq(h_1)}{h_1}.
\end{displaymath}
Notice that $LS=S\pq$, i.e., $(\pq+u_0+u_{1}\pq^{-1}+\cdots)(\pq+w_0)=(\pq+w_0)\pq,$
then the generalized dressing approach (Proposition \ref{prop:3})
 gives the one-soliton solution to the first type of $q$-KP equation with one source ((\ref{eqns:exam1}) with $N=1$)
\begin{eqnarray*}
\fl
  u_0=(1-\tht)(w_0)=(\tht-1)(\frac{\pq(h_1)}{h_1})\\
     =\frac{\lambda_1e_q(\lambda_1qx)e^{\xi_1}+\af_1(\ta_3)\mu_1e_q(\mu_1qx)e^{\et_1}}
     {e_q(\lambda_1qx)e^{\xi_1}+\af_1(\ta_3)e_q(\mu_1qx)e^{\et_1}}-\frac{\lambda_1e_q(\lambda_1x)e^{\xi_1}+\af_1(\ta_3)\mu_1e_q(\mu_1x)e^{\et_1}}
     {e_q(\lambda_1x)e^{\xi_1}+\af_1(\ta_3)e_q(\mu_1x)e^{\et_1}},
\end{eqnarray*}
\begin{eqnarray*}
\fl
  u_1=-[\pq(w_0)+(1-\tht)(w_0)w_0]=\frac{\partial_q^2 h_1}{h_1}-\left(\frac{\partial_q h_1}{h_1}\right)^2 \\
     =\frac{\lambda_1^2e_q(\lambda_1x)e^{\xi_1}+\af_1\mu_1^2e_q(\mu_1x)e^{\et_1}}
     {e_q(\lambda_1x)e^{\xi_1}+\af_1e_q(\mu_1x)e^{\et_1}}-
     \left(\frac{\lambda_1 e_q(\lambda_1x)e^{\xi_1}+\af_1\mu_1 e_q(\mu_1x)e^{\et_1}}
     {e_q(\lambda_1x)e^{\xi_1}+\af_1e_q(\mu_1x)e^{\et_1}}\right)^2,
\end{eqnarray*}
\begin{eqnarray*}
\fl
  u_2=-u_1\tht^{-1}(w_0)=u_1\theta^{-1}(\frac{\pq(h_1)}{h_1})
     =u_1 \frac{\lambda_1e_q(\lambda_1q^{-1}x)e^{\xi_1}+\af_1\mu_1e_q(\mu_1q^{-1}x)e^{\et_1}}
     {e_q(\lambda_1q^{-1}x)e^{\xi_1}+\af_1e_q(\mu_1q^{-1}x)e^{\et_1}},
\end{eqnarray*}
\begin{eqnarray*}
\fl
  \phi_1=-\dot{\af_1}\frac{h_1\pq(g_1)-\pq(h_1)g_1}{h_1}
    =-\frac{d \af_1}{d
    \tau_3}e_q(\mu_1x)e^{\et_1}[\mu_1-\frac{\lambda_1e_q(\lambda_1x)e^{\xi_1}+\af_1(\ta_3)\mu_1e_q(\mu_1x)e^{\et_1}}
     {e_q(\lambda_1x)e^{\xi_1}+\af_1(\ta_3)e_q(\mu_1x)e^{\et_1}}],
\end{eqnarray*}
\begin{eqnarray*}
\fl
  \psi_1=\tht(\frac1{h_1})=\frac1{e_q(\lambda_1qx)e^{\xi_1}+\af_1(\ta_3)e_q(\mu_1qx)e^{\et_1}}.
\end{eqnarray*}
\end{eg}

\begin{eg}
(One-soliton solution to the 1\it{st}-$q$-mKPSCS (\ref{meqn:exam1})) Let
$N=1$, then  by gauge transformation, the formulae (\ref{meqn:6}) gives
\begin{eqnarray*}
\fl
  \widetilde{L}=\widetilde{u}\pq+\widetilde{u}_0+\widetilde{u}_1\pq^{-1}+\cdots=(w_1\pq-1)\pq(w_1\pq-1)^{-1},\qquad w_1=\frac{h_1}{\pq(h_1)},
\end{eqnarray*}
this enables us to get the one-soliton solution to the first type of $q$-mKP equation with a source ((\ref{meqn:exam1}) with $N=1$)
\begin{eqnarray*}
\fl
  \widetilde{u}=\frac{w_1}{\tht(w_1)}=\frac{h_1\tht(\pq h_1)}{\pq(h_1)\tht(h_1)}\\
             =\frac{(e_q(\lambda_1x)e^{\xi_1}+\af_1e_q(\mu_1x)e^{\et_1})(\lambda_1e_q(\lambda_1qx)e^{\xi_1}+\af_1\mu_1e_q(\mu_1qx)e^{\et_1})}
                 {(\lambda_1e_q(\lambda_1x)e^{\xi_1}+\af_1(\ta_3)\mu_1e_q(\mu_1x)e^{\et_1})(e_q(\lambda_1qx)e^{\xi_1}+\af_1(\ta_3)e_q(\mu_1qx)e^{\et_1})},
\end{eqnarray*}
\begin{eqnarray*}
\fl
  \widetilde{u}_0=\frac1{w_1}[\widetilde{u}-1-\widetilde{u}\pq(w_1)]=\frac{\pq^2(h_1)}{\pq(h_1)}-\frac{\pq(h_1)}{h_1}\\
             =\frac{\lambda^2_1e_q(\lambda_1x)e^{\xi_1}+\af_1\mu^2_1e_q(\mu_1x)e^{\et_1}}
                 {\lambda_1e_q(\lambda_1 x)e^{\xi_1}+\af_1\mu_1e_q(\mu_1 x)e^{\et_1}}
                 -\frac{\lambda_1e_q(\lambda_1x)e^{\xi_1}+\af_1\mu_1e_q(\mu_1x)e^{\et_1}}
                 {e_q(\lambda_1x)e^{\xi_1}+\af_1e_q(\mu_1x)e^{\et_1}},
\end{eqnarray*}
\begin{eqnarray*}
\fl
  \widetilde{u}_1=\frac{\widetilde{u}_0}{\tht^{-1}(w_1)}=\widetilde{u}_0\frac{\tht^{-1}(\pq h_1)}{\tht^{-1}(h_1)}
             =\widetilde{u}_0\frac{\lambda_1e_q(\lambda_1q^{-1}x)e^{\xi_1}+\af_1\mu_1e_q(\mu_1q^{-1}x)e^{\et_1}}
                  {e_q(\lambda_1q^{-1}x)e^{\xi_1}+\af_1e_q(\mu_1q^{-1}x)e^{\et_1}},
\end{eqnarray*}
\begin{eqnarray*}
\fl
  \widetilde{\phi}_1=-\dot{\af_1}\frac{h_1\pq(g_1)-\pq(h_1)g_1}{\pq(h_1)}
            =-\frac{d\af_1}{d \tau_3}e_q(\mu_1x)e^{\et_1}[\mu_1\frac{e_q(\lambda_1x)e^{\xi_1}+\af_1e_q(\mu_1x)e^{\et_1}}
                 {\lambda_1e_q(\lambda_1x)e^{\xi_1}+\af_1\mu_1e_q(\mu_1x)e^{\et_1}}-1],
\end{eqnarray*}
\begin{eqnarray*}
\fl
  \widetilde{\psi}_1=\tht(\frac1{h_1})=\frac1{e_q(\lambda_1qx)e^{\xi_1}+\af_1e_q(\mu_1qx)e^{\et_1}}.
\end{eqnarray*}
\end{eg}

\section{Conclusions}
In this paper, we generalized the dressing approach for the $q$-KP
hierarchy to the $q$-KP hierarchy with self-consistent sources ($q$-KPHSCSs)
 by combining the dressing method and the method of variation of constants.
 The usual dressing method for the $q$-KP hierarchy can
not provide the evolution to the new $\tau_k$ variable. By
introducing some varying constants, say $\alpha(\tau_k)$, we can
obtain the desired evolution to $\tau_k$. In this way, we
constructed the $q$-deformed Wronskian solutions to
the $q$-KPHSCSs, and got the exact form for the
sources $\phi_i$'s and $\psi_i$'s.

On the base of eigenfunction
symmetry constraint, we constructed a $q$-mKP hierarchy with self-consistent sources ($q$-mKPHSCSs)
which contains two series of time variables, say $t_n$ and $\tau_k$.
The first and second type of $q$-mKP equation with sources ($q$-mKPSCS)  are obtained as the first
two non-trivial equations in the $q$-mKPHSCSs. And when $q\rightarrow
1$ and $\widetilde{u}\equiv 1$, the $q$-mKPHSCSs reduces to the
mKP hierarchy with self-consistent sources \cite{dress}.

A gauge transformation between the $q$-KPHSCSs and the $q$-mKPHSCSs is
established in this paper. By using the gauge transformation, we found the
Wronskian solutions for the $q$-mKPHSCSs. The one-soliton solutions to the $q$-KP equation with a source
((\ref{eqns:exam1}) with $N=1$) and to the $q$-mKP equation with a source
 ((\ref{meqn:exam1}) with $N=1$) are given explicitly.

{ 
It is interesting to consider if there exist solutions
in the $q$-deformed case, which are not surviving limit procedure to
the classical case. This will be studied in the future.}

\ack { 
The authors are grateful to the referees for the
quite valuable comments.} This work is supported by National Basic
Research Program of China (973 Program) (2007CB814800) and National
Natural Science Foundation of China (grand No. 10801083 and
10901090). RL acknowledges the economical support from ``Banco
Santander--Tsinghua University" program for his stay in UCM, and he
also thanks the Departamento de F\'{\i}sica Te\'{o}rica II (UCM) for
the warm hospitality.

\end{document}